\documentclass[11pt,twoside]{article}
\usepackage{epsfig}
\def\e{\begin{equation}}
\def\f{\end{equation}}
\def\=#1{\overline{\overline #1}}
\def\_#1{\overline #1}
\def\-#1{{\bf #1}}
\def\o{\omega}
\def\.{\cdot}

\def\l#1{\label{eq:#1}}
\def\r#1{(\ref{eq:#1})}
\def\am{\left(\begin{array}{c}}
\def\amm{\left(\begin{array}{cc}}
\def\a{\end{array}\right)}

\advance \textwidth by -5cm
\advance \textheight by -6cm
\begin{document}

\title{\bf{Toward creating wide-band uniaxial left-handed materials
with small losses}}

%%% Authors and affiliation - use for multiple authors

\def\affil#1{\begin{itemize} \item[] #1 \end{itemize}}

\author{C. R. Simovski}

\date{}
\maketitle \thispagestyle{empty}

\affil{Department of Physics, State Institute of Fine Mechanics and Optics\\
Sablinskaya Street, 14, 197101, St. Petersburg, Russia\\
Fax: 7-812-2322307; email: simovsky@phd.ifmo.ru}

\begin{abstract}
\noindent In this work a possible realizations of a uniaxial
variant of left-handed material (LHM) at microwaves is considered.
The meta-material has some features of the known structure studied
by the group of D. Smith in 2001, however in the present structure
a lattice of parallel wires and a lattice of artificial magnetic
resonators (MRs) are unified and the MRs are not split-ring
resonators. The optimizing of MRs allows significantly decrease
the magnetic losses and broaden the band within which the
meta-material becomes a LHM.
\end{abstract}

\subsection*{1.~Introduction}

Since 2000 the left-handed media (LHM) first introduced in
\cite{Veselago68} has become a subject of an abundant discussion
mainly due to J. Pendry \cite{lens}. In 2001, the negative
refraction in a prism of LHM was demonstrated at microwaves
\cite{Shelby2001}. This result has been reproduced in \cite{Lu}
and \cite{Houck} and now is considered by the scientific community
as a reliable one. Negative real part of effective permittivity
$\epsilon_{eff}$ in the two-phase composite studied by D. Smith,
R. Shelby and S. Schultz in their work \cite{Shelby2001} was
created by an array of parallel conducting wires (the wire axis
$x$ is the optical axis of this composite medium). The negative
value of the real part of permeability $\mu_{eff}$ was provided by
artificial magnetic resonators (MRs) performed as double
split-ring resonators (SRRs). The known SRRs are two coplanar
\cite{Pendry1999} or parallel \cite{Marques} split wire rings. The
negative permeability arises due to the resonance of the magnetic
polarizability of SRRs {within a} narrow sub-band which belongs to
their resonant band. This sub-band lies in the wide frequency
range where ${\rm Re}(\epsilon_{eff})<0$ (more exactly this is the
real part of the $xx-$component of the permittivity tensor). As a
result, the material becomes a uniaxially anisotropic LHM within
this sub-band. Electromagnetic waves in this LHM suffer the
magnetic and electric losses. There has been many papers written
about these losses since the work \cite{Garcia} in which the
structure tested in \cite{Shelby2001} was treated as very lossy
and claimed therefore not to be a LHM. Comparing the results
obtained in \cite{Garcia}--\cite{Pokrovsky} with the results of
\cite{Soukoulis} we can see that different simulations of the same
structure (Smith-Shelby lattice) give very different results for
its losses. Experimental data do not fit with all these
simulations and give moderate values for transmission losses per
unit thickness in the LHM regime (see \cite{Houck} and \cite{Lu}).
In the paper \cite{new} the influence of the electromagnetic
interaction of SRRs with wires to the losses in this medium has
been explaned. One has obtained for ${\rm Im}(\mu_{eff})$ the
values of the order $0.1{\rm Re}(\mu_{eff})$ within the band where
${\rm Re}(\mu_{eff})<0,\ {\rm Re}(\epsilon_{eff})<0$. It is clear,
that the losses are rather significant in this structure. These
losses can be slightly reduced increasing the thickness of wires
from which SRRs are fabricated, but this increase is restricted by
the need to have the strong magnetic resonance. Another short of
this structure is a narrow frequency band in which the
meta-material is a LHM (in \cite{Shelby2001} this band was equal
few tens of MHz centered at $4.5$ GHz).

In the literature one can find alternative versions of a LHM at
microwaves, e.g. \cite{sim1} and \cite{book}. In \cite{sim1} the
self-consistent analytical theory of the quasi-isotropic lattice
of metal bianisotropic (Omega) particles has been presented, and
the negative parameters predicted within the rather large band
$8.2\dots 8.4$ GHz. However, the losses were neglected in this
work. In \cite{book} the experimental testing of a LHM made from
SRRs combined with capacitively loaded strips (instead of long
wires) has been done. The aim of \cite{book} was to match this
medium with free space. The measured losses turned out to be very
high \footnote{The mini-pass-band which theoretically
corresponding to negative material parameters is almost invisible
within the band of the resonant absorbtion of SRRs, judging on the
plots of the transmittance for a $290$ mil-thick ($7.25$ mm)
layer. This plot is shown in Figs. 10 and 14a.}, and the band in
which both material parameters extracted from measured data have
negative real parts is rather narrow. In Fig. 14 (b) it can be
detected as $50$ MHz (centered at $9.5$ GHz\footnote{In Fig. 17
there is another band of negative $\epsilon_{eff},\mu_{eff}$
($10.9-11.1$ GHz), however at these frequencies the lattice period
becomes larger than $\lambda/4$ and the usual constitutive
parameters are not physically sound.}). In \cite{proceedings} the
transmittance through the layer of a racemic medium from resonant
chiral particles was calculated. This medium also exhibits
negative real part of constitutive parameters within the resonant
band of particles. The result for the resonant losses was also
pessimistic \cite{proceedings}.

\begin{figure}
\centering \epsfig{file=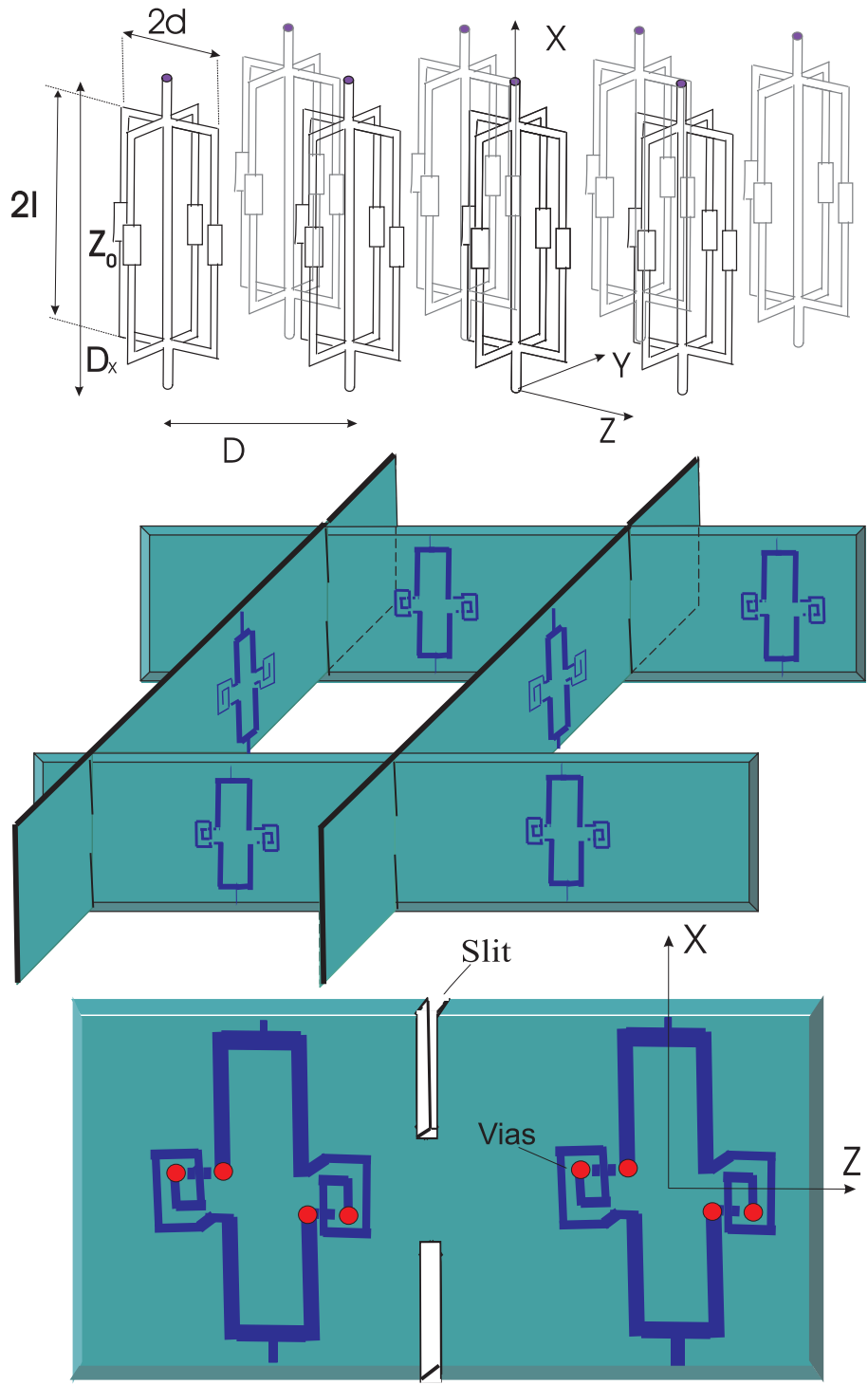, width=12cm}
 \caption {Structure under study (on top) and the idea of its planar
 realization (on bottom). Inductive loads are performed as planar spirals
 with conducting vias.} \label{fig1}
\end{figure}

In the present paper we suggest another uniaxial variant of LHM
structure. The structure is drawn in Fig. \ref{fig1}, on top. The
model developed below describes an infinite square lattice
prepared from wires located in a homogeneous dielectric host
medium of permittivity $\epsilon_m$. Note, that the properties of
the similar structure prepared with the use of planar technology
and shown on bottom of Fig. \ref{fig1} are expected to be similar.
This must be so if the lattice period $D$ is small enough compared
with the wavelength. Then the dielectric board is practically a
mixture and its averaged response to the field must be close to
that of the host medium with certain effective permittivity
$\epsilon_m$ (lying between that of free space and that of the
dielectric plates). Though a metal strip of width $w$ possess more
significant ohmic losses per unit length than a wire with round
cross section of same diameter $w$, the frequency dependence of
these losses is approximately the same in both cases. Therefore
one can practically find an equivalent radius of the round wire
$r_0$ for given $w$. Replacing the host medium by the board of
dielectric plates and the round wires by the strip wires we
theoretically do not change the low-frequency properties of the
structure. This makes the variant of LHM with loaded loops shown
in Fig. \ref{fig1} realizable in scientific laboratories. The
procedure of obtaining such lattices (using vertical slits) is
rather simple. It has been described in \cite{mybook} and recently
used by Dr. Sauviac from the laboratory DIOM (University of St.
Etienne, France) for preparing the prototype of the left-handed
composite introduced in \cite{sim1} (its preliminary experimental
study is presented in \cite{new2}). The main difference between
structures drawn on top and on bottom of Fig. \ref{fig1} is that
the structure on top contains one straight wire in a unit cell and
the structure on bottom contains two wires. This difference is not
crucial.

The MRs suggested in the present paper are long loops (their
length $2l$ is of the order $\lambda/2$). It is also assumed that
$2l$ is rather close to $D_x$ (the period of loops along the
$x-$axis). Each loop contains two symmetrically located equivalent
lumped loads. The width of the loop $2d$ is small ($2d\ll D\ll
\lambda$). In Fig. \ref{fig1} we have shown the case of an
inductive load $Z_0=j\o L_0$ performed as a small planar spiral (a
conventional lumped inductance at microwaves). Linear chains of
MRs are electrically connected with the short wires. These line
chains operate as straight thick wires being excited by the
electric field and as linear arrays of MRs being excited by the
magnetic field. Unifying the wires and MRs as it is shown in the
Figure allows increase the effective radius of wires and reduce
the ohmic losses. The suggested geometry of a MR helps to obtain a
rather wide band for an effective LHM (below an example with $120$
MHz at $7$ GHz is given) in which the $|{\rm Im}\mu_{eff}|\ll
0.1|{\rm Re}\mu_{eff}|$. Imaginary part of $\epsilon_{eff}$ is
always smaller than $|{\rm Im}\mu_{eff}|$.

\subsection*{2.~Dispersion equation}

Let us consider a square lattice with step $D_y=D_z=D$ of chains
presented in Fig. \ref{fig1}, on top. Let the wave with unknown
propagation constant $\beta$ (that of the main Bloch mode)
propagates along $z$. To obtain a dispersion equation of the
lattice one can consider the response of the chain to the
$\-E-$field and $\-H-$field averaged over the plane $(x-y)$ and
directed along $x$ and $y$, respectively. Let us denote these
averaged values as $<E>$ and $<H>$, respectively. The scale of
averaging along $x$ is $D_x$, and along $y$ it is $D$. The
electric field $<E>$ excites two orthogonal loops stretched along
the $x-$axis (see Fig. \ref{fig1}) as a thick wire because the
induced currents will be parallel in all four sides of the two
orthogonal loops (at $x=\pm d,\ y=0$ and $y=\pm d,\ x=0$). To this
equivalent thick wire an effective radius $\rho$ can be
attributed, and it is clear that $r_0<\rho<d$, where $r_0$ is the
radius of the real wire cross section. So, the chain excited by
field $<E>$ operates as a wire with radius periodically changing
along $x$ from real wire radius to $\rho$. This periodicity of the
radius along $x$ has no impact to the waves propagating
orthogonally to $x$ and we can approximately consider our chain
excited by electric field as a wire with effective uniform radius
$r$. In our numerical example below we have chosen $r=d/10=0.1$
mm, whereas the radius $r_0$ of the wire forming the loops has
been assumed to be equal $r_0=0.02$ mm. We do not know the exact
relations between $r_0$ and $r$, however the ratio $r/r_0=5$ can
be, of course, realized in practice. This is so, because there is
a free parameter for tuning the equivalent radius $r$: the radius
of the wire connecting two adjacent loops. We will see below that
the value of $r_0$ is important for calculating the ohmic losses
in the MRs and (as a result) for magnetic losses in the lattice.

We conclude that the electric excitation of the array of parallel
chains located in the plane $(x-y)$ (see Fig. \ref{fig1}) is
equivalent to the excitation of a mesh of parallel wires with
radius $r$ and period $D_y=D$. The grid impedance $Z_g$ of the
wire mesh is well-known. It relates the averaged surface current
$J$ of the mesh with $<E>$ and can be found from the following
relations (see e.g. in \cite{new1}): \e <E>=Z_gJ={j\eta\alpha\over
2}J,\qquad \alpha= {kD_y\over \pi}\log {D_y\over 2\pi r}.
\l{grid}\f Here $\eta=\sqrt{\mu_0/ \epsilon_0\epsilon_m}$ is the
wave impedance of the host medium. This approach allows to
consider an array of chains excited by electric field as a sheet
of electric current $J$. Then the lattice in which the chains are
excited by electric field can be treated as a set of parallel
sheets of electric current.

Relations \r{grid} are accurate enough for dense wire grids
($kD<1$) and are valid for the lossless case. In the present paper
we compare the result obtained for our structure with results for
the lattice of wires and SRRs obtained analytically in our work
\cite{new}. In the cited work we took into account the finite
conductivity of straight wires but it practically did not change
the result compared to the case of the perfect wires. In
accordance with \cite{new} the wave attenuation is practically due
to the finite conductivity of SRRs. Therefore in our comparison
with \cite{new} we will neglect the lossy part of $Z_g$ but we
will take into account the ohmic losses in MRs.

Now let the same array of chains in the $(x-y)$ plane be excited
by field $<H>$. First, notice that the loops of the reference
chain (see Fig. \ref{fig1}) lying in this plane are not excited by
$y$-directed magnetic field. The chain is excited due to the loops
lying in the $(x-z)$ plane. Using Maxwell's equation
$$
-j\o\mu_0 <H>={d <E>\over d z}\approx {1\over 2d
}\left(<E>(z=d)-<E>(z=-d)\right)
$$
we easily obtain that the magnetic excitation of the chain is the
same as if the wire located at $z=d$ was excited by $x-$directed
electric field $E=j\o\mu_0 d<H>$, and the wire located at $z=-d/2$
was excited by electric field $E=-j\o\mu_0 d<H>$.

Then we can find the effective voltage ${\cal E}$ exciting the
loop using the method of induced electromotive forces:
$$
{\cal E}=\int_{L} <E>(\tau) f(\tau) d\tau,
$$
where $f(\tau)$ is the current distribution along the contour $L$
of our loop, normalized to the induced current $I_0$ taken at the
point to which the voltage ${\cal E}$ is referred. Let us refer
${\cal E}$ to the center of the loop (to a point of the lumped
load). Then this voltage is connected to two shortened lines of
length $l$ in series with two lumped loads $Z_0$. The current
distribution in the shortened line of length $l$ is approximately
sinusoidal and has the maximum at the shortcut point ($x=\pm l$):
$$
I(x)=I_0f(x)=I_0\cos k(x-l)/cos kl.
$$
Here $k=\o\sqrt{\epsilon_0\epsilon_m\mu_0}$ is the wave number of
the host medium. Then we obtain \e {\cal E}=-4j\o\mu_0 {\tan
kl\over k} d <H>. \l{emf}\f If $kl\ll 1$ formula \r{emf} transits
to the Faraday formula ${\cal E}=-j\o\mu_0 S <H>$, where $S=4d l$
is the area of the loop. The impedance to which the voltage ${\cal
E}$ is connected is \e Z=2(R+jW \tan kl + Z_0)=2R+2j(W \tan kl +
X_0), \l{zz}\f where $X_0$ is the loading reactance, $R$ is the
ohmic resistance of the line, $W$ is its wave impedance
\cite{Gardiol}:
$$
R=\sqrt{\mu_0\o\over 2\pi^2\sigma}{l\over
r_0\sqrt{1-\left({2r_0\over d}\right)^2}},\quad W={\eta\log{d\over
r_0}\over \pi}.
$$
In \r{zz} we neglected the radiation resistance of the loop,
however, it does not lead to a mistake because in regular arrays
the radiation resistance is suppressed by the interaction of the
array elements \cite{JOSA}.

We find the current as $I_0={\cal E}/Z$ and then find the magnetic
moment per unit length of a chain defined as $m=2d\mu_0<I>$, where
$<I>$ is the averaged (along $x$) value of the current $I$:
$$
<I>={2I_0\over D_x}\int\limits_{0}^l f(x) dx.
$$
The magnetic moment per unit surface of the plane $(x-y)$ is
related with $m$ as $M=m/D_y=2d\mu_0<I>/D$. Then \e
M={2d\mu_0{\cal E} \over DZ}. \l{mm}\f Below we use the magnetic
susceptibility of the reference sheet (located in the plane
$(y-z)$) defined as $a_{mm}=M/<H>$. Substituting \r{emf} and
\r{zz} into \r{mm} we obtain for this susceptibility the relation
\e {1\over a_{mm}}=\left[-(W\tan kl+X_0)+jR\right] \left({kl\over
\mu_0S\tan kl}\right)^2{S_0\over \o}. \l{fin}\f Here $S_0$ is the
area of the mesh unit cell $S_0=D_xD$. If the load is capacitive
$X_0=1/\o C_0$ and the loop is electrically small $kl\ll 1$
formula \r{fin} gives a well-known result for a split-ring
resonator (see \cite{Pendry1999}):
$$
a_{mm}=G{\o^2\over {1\over L_{\rm loop}C_0}-\o^2+j{R\over L_{\rm
loop}}\o},
$$
where $G$ and $L_{loop}$  are constants and the constant $L_{\rm
loop}=Wl\sqrt{\epsilon_0\epsilon_m\mu_0}$ plays role of the
effective inductance of the long narrow loop.

Our lattice excited by an electromagnetic wave is a set of
parallel magneto-dielectric sheets with electric susceptibility
$a_{ee}=1/Z_g$ and magnetic one $a_{mm}$. These susceptibilities
are defined as \e J=a_{ee}<E>,\quad M=a_{mm}<H>. \l{def}\f Each
sheet produces a plane wave. Let us enumerate the sheets so that
the reference one has number $n_z=0$ (then $M=M(n_z=0)$ and
$J=J(n_z=0)$). Since $M(n_z)$ and $J(n_z)$ satisfy
$$
M(n_z)=M(0)e^{-jn_z\beta D},\quad J(n_z)=J(0)e^{-jn_z\beta D},
$$
we easily obtain for the $\-E$-field and $\-H$-field in the
reference plane the following set of relations:
$$
<E>=<E>^{J}+<E>^{M},\quad <H>=<H>^{J}+<H>^{M},
$$
$$
<E>^{M}={j\o M\over 2}\sum_{-\infty}^{\infty}{n_z\over
|n_z|}e^{-jn_z\beta D-jk|n_z|D},
$$
$$
<H>^{J}={ J\over 2}\sum_{-\infty}^{\infty}{n_z\over
|n_z|}e^{-jn_z\beta D-jk|n_z|D},
$$
$$
<E>^{J}=-{\eta J\over 2}\sum_{-\infty}^{\infty}e^{-jn_z\beta
D-jk|n_z|D}
$$
and
$$
<H>^{M}={j\o M\over 2\eta}\sum_{-\infty}^{\infty}e^{-jn_z\beta
D-jk|n_z|D}.
$$
Terms $<E^{M}>,\ <H>^{J}$ describe the electro-magnetic and
magneto-electric interaction of the sheets. Terms $<E^{J}>,\
<H>^{M}$ describe electro--electric and magneto-magnetic
interaction. All these series can be easily summarized, and we
obtain \e <E>= \kappa_{ee} AJ+ \kappa_{em} BM \l{eemm}\f and
 \e <H>=
\kappa_{mm} AM+ \kappa_{me} BM, \l{emme}\f where it is denoted \e
\kappa_{em}=-{\o\over 2}={\kappa_{me}\over j\o},\qquad
\kappa_{ee}={-j\eta\over 2}={\kappa_{mm}\over j\o\eta^2}
\l{kappa}\f and \e A={\sin kD\over \cos kD -\cos\beta D}, \quad
B={\sin \beta D\over \cos kD -\cos\beta D}. \l{ab}\f Substituting
\r{eemm} and \r{emme} into \r{def} we obtain a set \e
(1-a_{mm}\kappa_{mm}A) M-a_{mm}\kappa_{me}B J=0, \l{first}\f \e
(1-a_{ee}\kappa_{ee}A) J-a_{ee}\kappa_{em}B M=0, \l{second}\f
which immediately leads to the dispersion equation: \e
\left({1\over a_{mm}}-\kappa_{mm}A \right) \left({1\over
a_{ee}}-\kappa_{ee}A \right)=\kappa_{me}\kappa_{em}B^2.
 \l{disp}\f
Taking into account relations \r{kappa}, \r{ab} and \r{grid}
(which can be written in form $1/a_{ee}=Z_g=j\eta\alpha/2$) one
can rewrite \r{disp} in a following form: \e {2\over \o\eta
a_{mm}}(\cos kD -\cos\beta D)\sin kD- (1+\alpha)\sin^2
kD-\cos^2\beta D+1=0.
 \l{disp1}\f
This is the needed dispersion equation of our structure. Notice,
that for the lossless case, we have obtained the real dispersion
equation providing the band structure of our lattice. This
equation is valid not only at low frequencies ($kD<1$), where the
grid parameter $\alpha$ is given by \r{grid}, it is valid at
rather high frequencies where the high-order corrections to
$\alpha$ can be taken into account (see e.g. in \cite{new1}).
Equation \r{disp1} has two solutions. One solution describes the
propagation factor of the wave which interacts with the lattice
and is polarized so that $E=E_x,\ H=H_y$. Another solution is
trivial ($\beta =k$) and corresponds to the case $E=E_y,\ H=H_x$.

\begin{figure}
\centering \epsfig{file=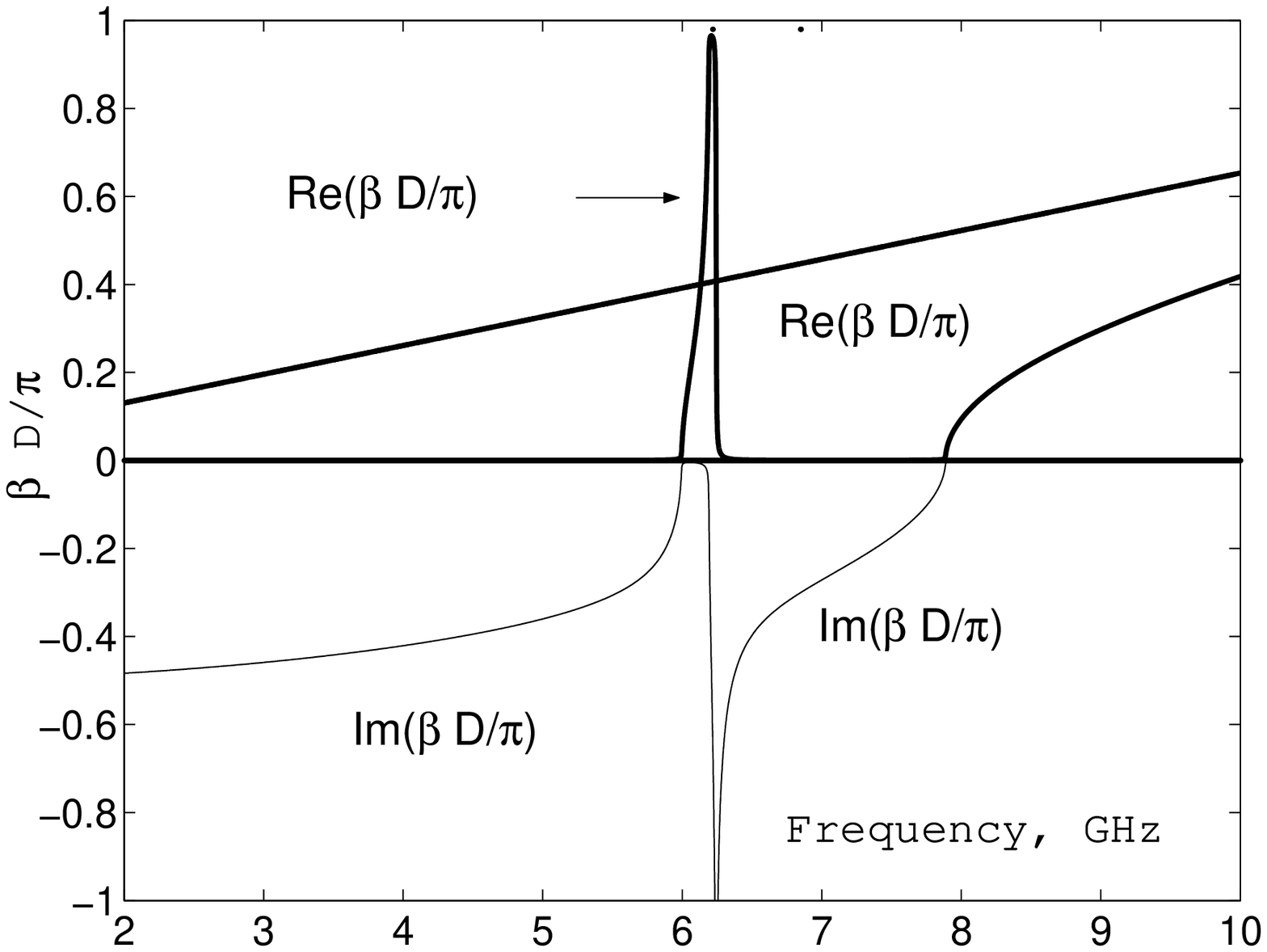, width=7.5cm}
\epsfig{file=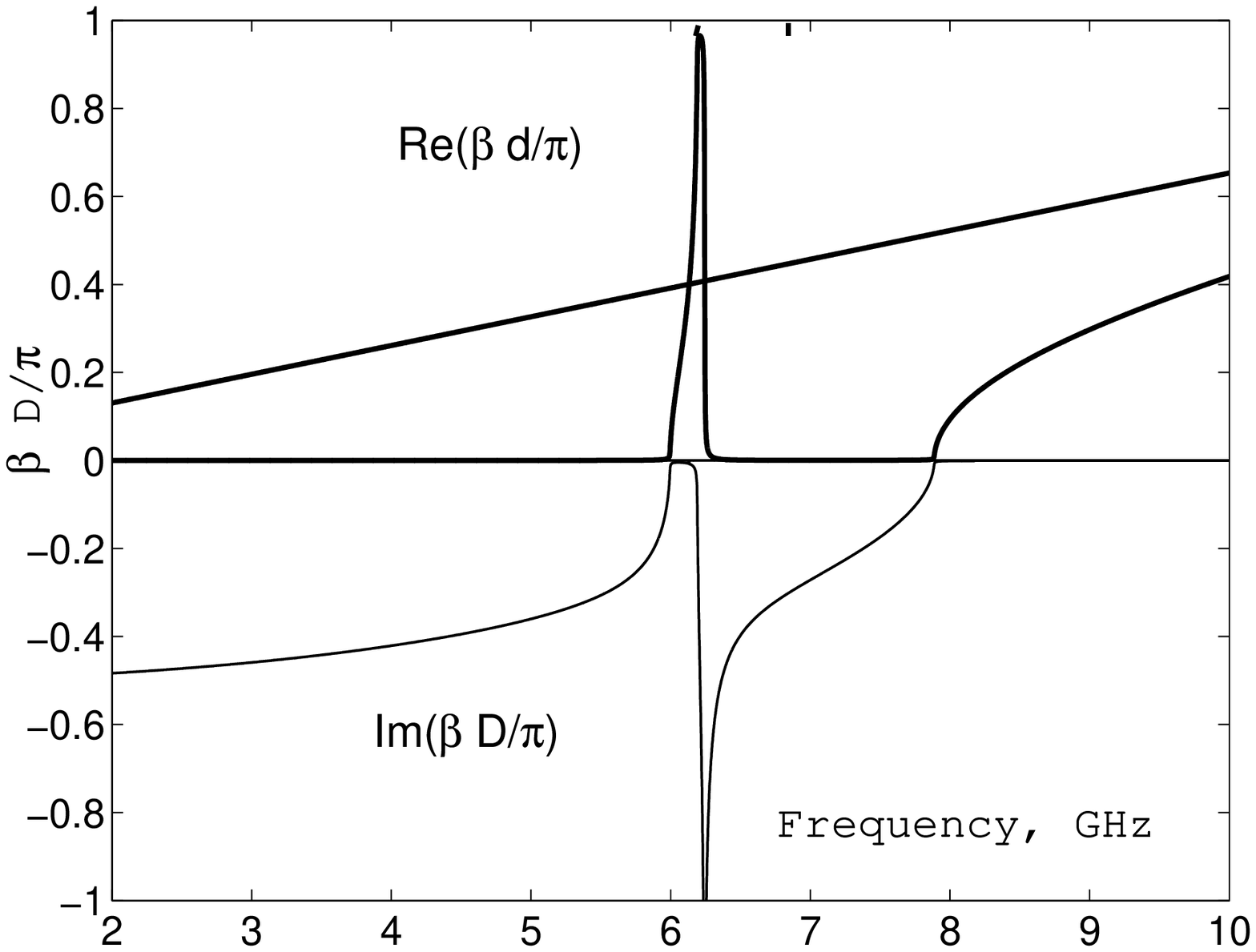, width=7.5cm}
 \caption {Dispersion curves for our structure without loop
 loading. Left: perfect wires. Right: copper wires. Straight lines correspond to
 the trivial solution of the dispersion equation.} \label{fig2}
\end{figure}

The results for the normalized propagation factor versus frequency
(in GHz) are shown in Fig. \ref{fig2}. The following parameters of
the lattice were picked up: $2d=2$ mm, $D=8$ mm, $D_x=24$ mm,
$l=10$ mm, $Z_0=0$ (no loop loading) and $r_0=0.02$ mm. The plot
at the left of this Figure corresponds to perfectly conducting
chains. The plot at the right corresponds to the chains fabricated
from copper. The lattice is assumed to be located in the
homogeneous dielectric matrix with permittivity
$\epsilon_m=1.5-j0.001$. The losses of the matrix do not play any
visible role, and the dispersion plot at the left is practically
the same as for the lossless dielectric.

The resonant band of MRs contains two sub-bands. In the upper half
of the resonant band ($6.21-6.42$ GHz) the lossless lattice has
the complex mode $\beta D=\pi+j{\rm Im}(\beta D)$. This is a
well-known mode of metallic photonic crystals. The propagation is
suppressed whereas the induced currents are alternating from one
chain to another along the $z-$axis. In the lower half of the
resonant band ($6.00-6.21$ GHz) the lossless lattice supports the
usual forward wave, and below we will see that this is the domain
where $\epsilon_{xx}>0$ and $\mu_{yy}>0$. This difference with the
known case of SRRs and wires corresponds to the substitution of a
series resonance (of SRRs) by a parallel one (of our MRs).

The ohmic losses in MRs visibly shift the resonance. This effect
was explained in \cite{new} (for the case when MRs are performed
as SRRs) as the result of electromagnetic interaction in the
lattice. In the lossy case the complex mode disappears. The lower
sub-band still corresponds to the usual (forward-wave) mode
propagating with negligible attenuation, whereas the backward wave
in the upper sub-band suffers very strong attenuation. Below we
will see that this sub-band corresponds to the case ${\rm
Re}(\epsilon_{xx})<0$ and ${\rm Re}(\mu_{yy})>0$.

However one can significantly decrease this attenuation
substituting the parallel resonance with the series one. For it we
introduce the loading of the loops.  The loaded loop can have in
the same frequency range both series and parallel resonances. The
capacitive loading $Z_0=1/j\o C_0$ gives the series resonance as a
first resonance versus frequency. At this resonance our MRs
operate as SRRs, and the attenuation of the backward wave turns
out to be rather important. No advantages was found for this case
in the replacing the SRR by long loops.

\begin{figure}
\centering \epsfig{file=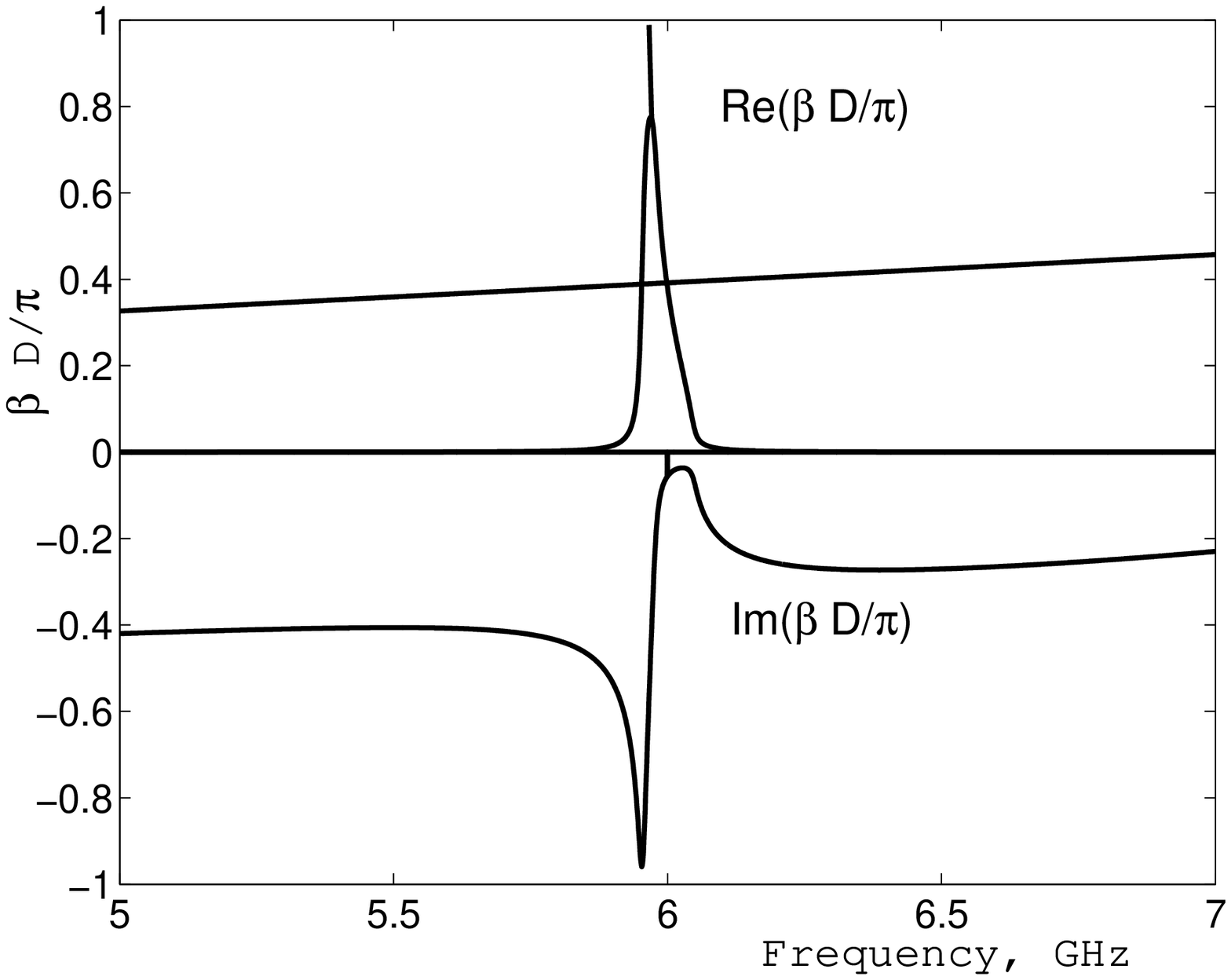, width=7.5cm}
\epsfig{file=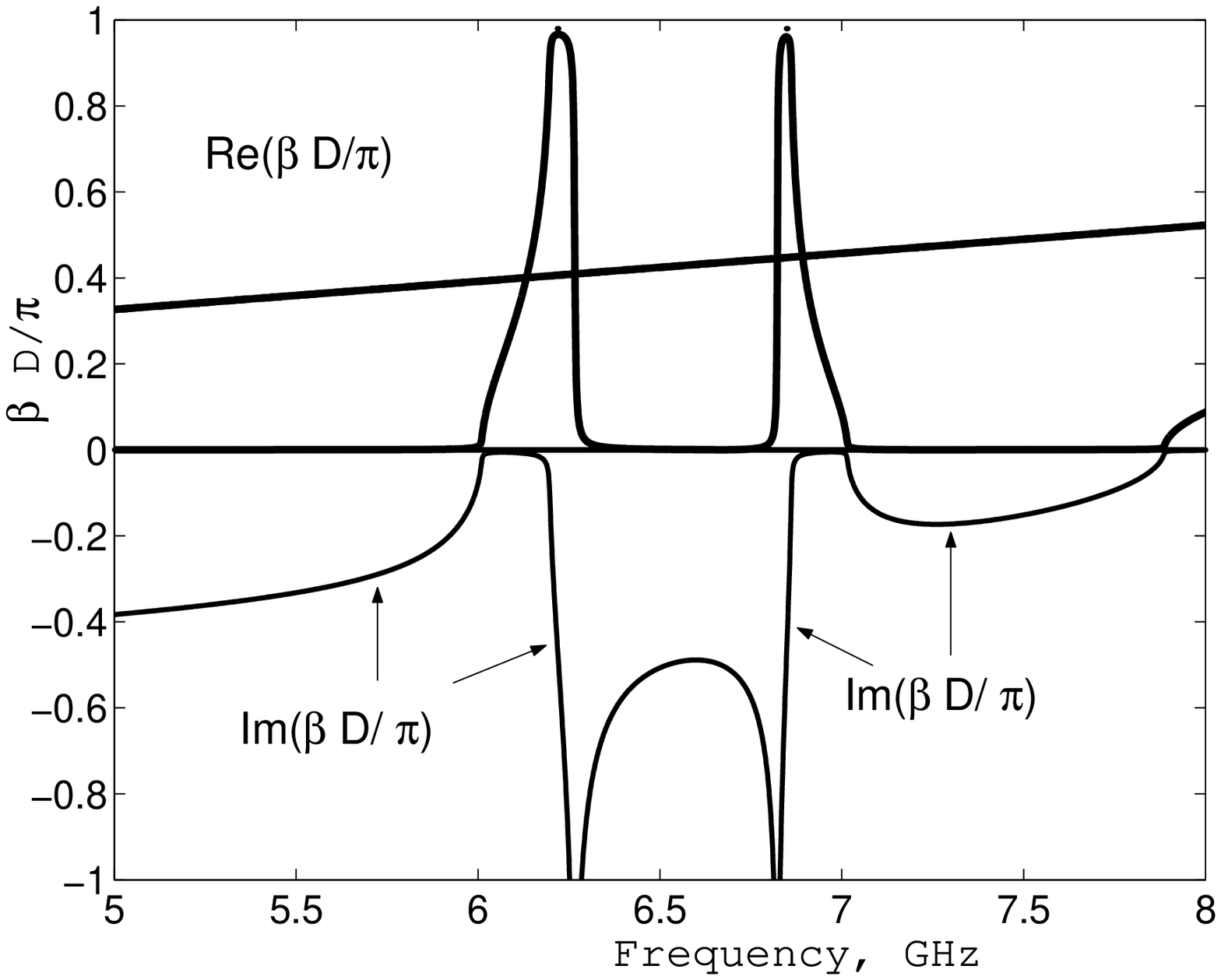, width=7.5cm}
 \caption {Dispersion curves for two left-handed media from copper wires.
Left: the lattice of straight wires and SRRs. Right: our structure
with inductive loads. Straight lines correspond to the trivial
solutions of dispersion equations.} \label{fig3}
\end{figure}

The inductive loading $Z_0=j\o L_0$ is more useful. It not only
shifts the frequencies of the series and parallel resonances of
our loop, it also broadens the resonant band. Moreover, the
attenuation of the backward wave can be strongly decreased by the
appropriate choice of $L_0$. In Fig. \ref{fig3}, right part, we
show the dispersion plot for the case when $L_0=30$ nH (it is a
realistic value for a microwave lumped inductance of size $1$ mm).
Other parameters are the same, but the new length of the loop is
chosen $2l=17$ mm to keep nearly the same frequency range of the
resonance. Loaded MRs have now two resonance bands at the
frequencies from $0$ to $10$ GHz. The lower one is still the
antiresonance, and the second one is the series resonance. Within
both of them there is a sub-band of the backward wave. In the
first resonance sub-band the attenuation of this backward wave is
very high (this effect will be discussed below). In the second
sub-band of the backward wave (near $7$ GHz) the attenuation is
small (almost invisible in the plot). For comparison the
dispersion plot of the two-phase lattice from wires and SRRs is
presented in the same Figure (left part). There the attenuation of
the backward wave (within $6-6.1$ GHz) is visible clearly. The
period of both wire and SRR lattices was taken the same as for our
lattice ($D=8$ mm), as well as the permittivity of the matrix
($\epsilon_m=1.5-j0.001$). The parameters of SRRs were picked up
so that to obtain the resonance at nearly the same frequencies:
$r_1=1.5$ mm (radius of the outer ring), $r_2=1.2$ mm (that of the
inner ring), wire radius $r_0=0.05$ mm. The SRRs also were assumed
to be made from copper. The calculations corresponds to the
analytical model described in \cite{new}. It is important that the
structure suggested in the present paper has smaller attenuation
of the backward wave than the lattice of SRRs and straight wires,
though the rather large radius $r_0=50\ \mu$m was picked up for
the two-phase lattice (versus $r_0=20\ \mu$m picked up for our
structure). One can see in the same Figure that the model also
predicts for our structure more wide frequency band of the
backward wave than for the lattice of SRRs and wires.

\subsection*{3.~Material parameters}

let us try to consider the structure as a uniaxial
magneto-dielectric medium. Then the interacting wave satisfies the
following material equations: \e {\cal
D}_x=\epsilon_0\epsilon_mE_x+P^{\rm
bulk}_x=\epsilon_0\epsilon_{xx}E_x,
 \l{d}\f
\e {\cal B}_y=\mu_0H_y+M^{\rm bulk}_y=\mu_0\mu_{yy}H_y.
 \l{b}\f
Let us find $\epsilon_{xx}$ and $\mu_{yy}$. Define the auxiliary
parameter $\gamma$ as
$$
\gamma=\eta{P^{\rm bulk}_x\over M^{\rm
bulk}_y}={(\epsilon_{xx}-1)E_x\over \eta(\mu_{yy}-1)H_y}.
$$
>From our definitions of $J$ and $M$ it follows that
$$
P^{\rm bulk}_x={J\over j\o D_y}={J\over j\o D},\quad M^{\rm
bulk}_y={M\over D_z}={M\over D}.
$$
>From Maxwell's equations together with \r{d} and \r{b} one has:
$$
{E_x\over H_y}=\eta\sqrt{\mu_{yy}\over \epsilon_{xx}}
$$
Therefore \e \gamma={\eta J\over j\o M}={(\epsilon_{xx}-1)\over
(\mu_{yy}-1)} \sqrt{\mu_{yy}\over \epsilon_{xx}}. \l{final}\f From
\r{first} or \r{second} we easily find $J/M$. Then  \r{final}
leads to the first equation for material parameters: \e
{(\epsilon_{xx}-1)\over (\mu_{yy}-1)} \sqrt{\mu_{yy}\over
\epsilon_{xx}}={\eta(1-a_{mm}\kappa_{mm}A)\over j\o
a_{mm}\kappa_{me}}. \l{fir}\f For $A(\beta)$ we have the relation
\r{ab},and $\beta$ is found from dispersion equation \r{disp1}.
The second equation is trivial \e
\beta^2=k^2\epsilon_{xx}\mu_{yy}.
 \l{sec}\f
>From \r{fir} and \r{sec} we find two solutions for $\epsilon$ and
$\mu$ and pick up the true one having the negative imaginary part.

\begin{figure}
\centering \epsfig{file=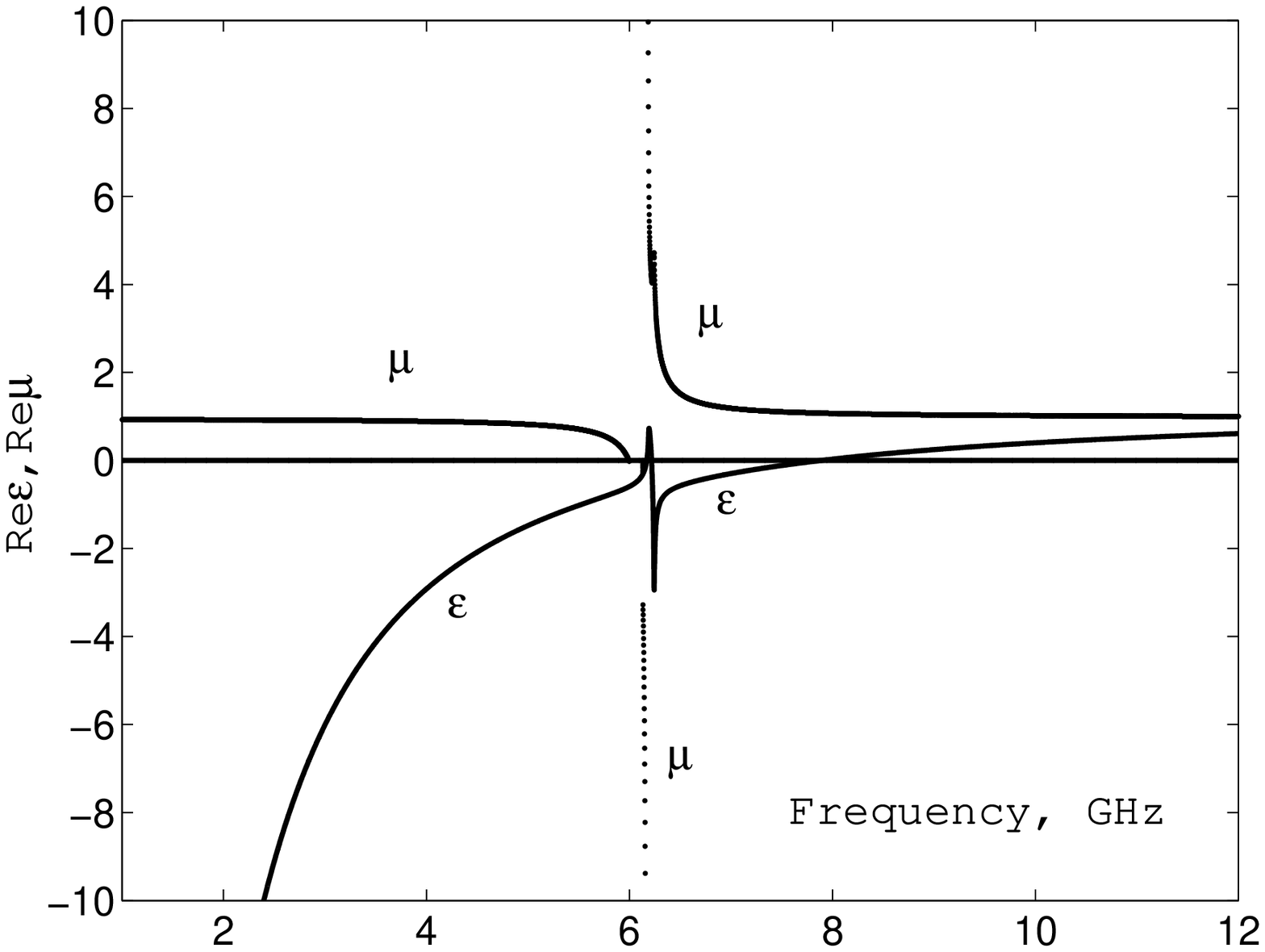, width=7.5cm}
\epsfig{file=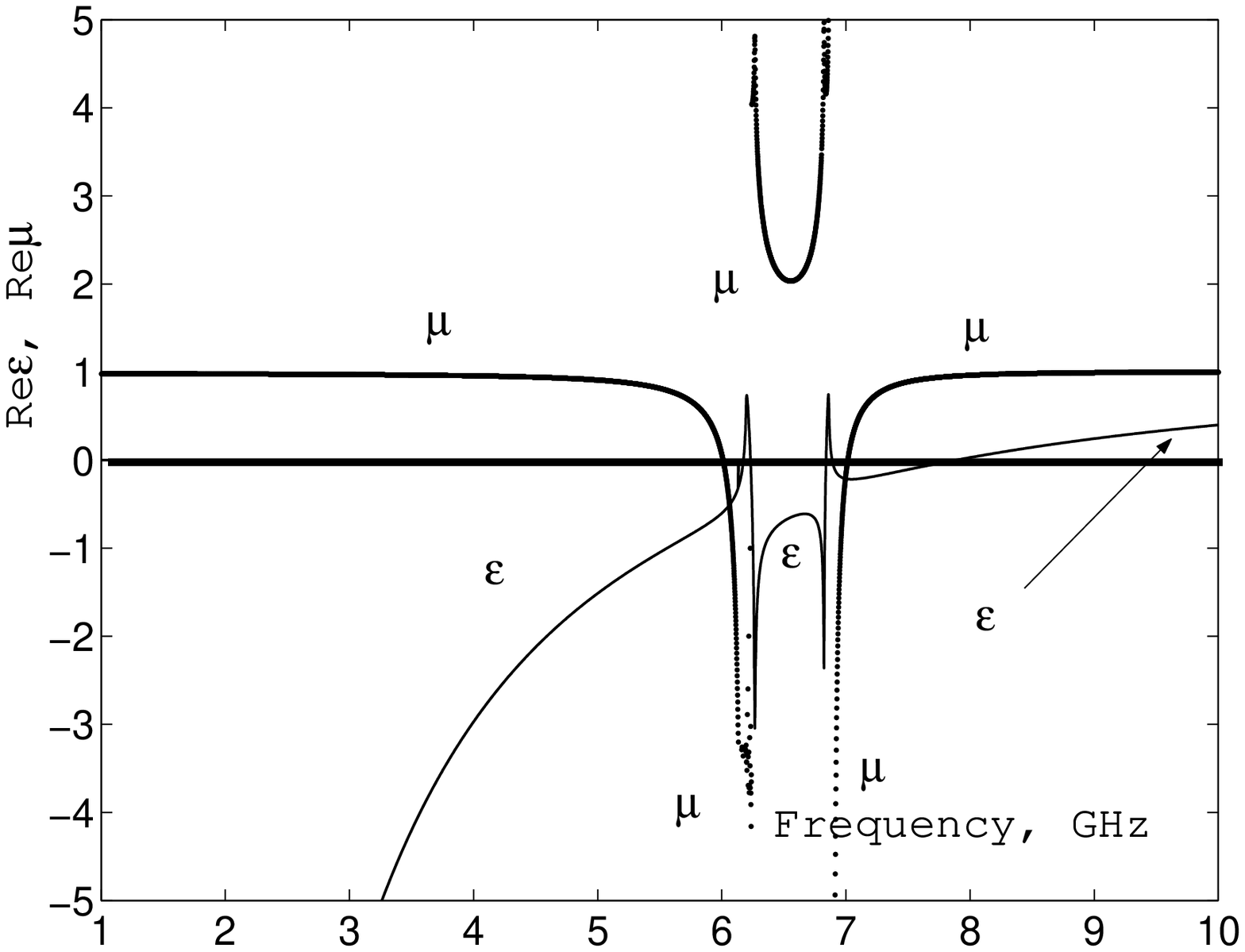, width=7.5cm}
 \caption {Effective constitutive parameters
$\epsilon_{xx}$ and $\mu_{yy}$ (real parts only) of the lattice
from chains of long loops. Left: unloaded loops. Right: loaded
loops.} \label{fig4}
\end{figure}

In Fig. \ref{fig4} we present the real parts of permittivity
(solid lines) and permeability (dotted lines). At the left the
case of unloaded loops is shown (corresponding to the right part
of Fig. \ref{fig2}). At the right the case of loaded loops is
presented (corresponding to the right part of Fig. \ref{fig3}).
The resonant behavior of permittivity is an illustration of the
fact, that a naive understanding of permittivity and permeability
of left-handed composites as an arithmetic sum of the permittivity
of wire medium and permeability of an artificial magnetic is
completely wrong. The electromagnetic interaction of magnetic and
electric components of the meta-material tightly relates these two
parameters.

One can see that the case without loads (parallel resonance at $6$
GHz) corresponds to the case when $\mu$ and $\epsilon$ are never
have negative real parts simultaneously\footnote{Since there is no
load the first series resonance of the loop corresponds to $12$
GHz. At this resonance the material parameters formally become
both negative but loose the physical meaning, since the lattice
period $8$ mm becomes larger $\lambda/4$.}. The lower sub-band of
the parallel resonance band corresponds to the very small
attenuation and contains two very narrow frequency bands. In the
lower ($6-6.1$ GHz) one the permeability becomes negative, but the
permittivity simultaneously becomes positive. This is still the
stop-band in Fig. \ref{fig2} (the right part of this Figure
corresponds to the copper wires). In the upper one ($6.1-6.2$ GHz)
the permeability becomes positive, whereas the resonant
permittivity is still positive. This is the band of the forward
wave shown in Fig. \ref{fig2}. The frequencies $6.2-6.24$ GHz are
the upper sub-band of the resonant band in which there appears a
strong attenuation, the imaginary parts of material parameters
become very large with respect to the real parts, and ${\rm
Re}(\epsilon_{xx})<0$, whereas ${\rm Re}(\mu_{yy})>0$. This is the
band of the attenuating backward wave shown in Fig. \ref{fig2},
right part (in the lossless case, presented in Fig. \ref{fig2} at
the left, it is not a backward wave but a complex mode).

Backward wave shown in the right parts of Figs. \ref{fig2} and
\ref{fig3} which (in the lossy case) replaces the complex mode of
a lossless lattice and which corresponds to the parallel resonance
of MRs has nothing to do with left-handed medium. When the
imaginary parts of material parameters are larger than their real
parts the group velocity has no physical meaning and the negative
refraction cannot be observed for real signals. Moreover, this is
not a continuous medium at all. Huge values of ${\rm
Im}(\mu_{yy})$ mean that the characteristic scale of the field
variation is smaller than the period, and the material parameters
we calculated for this sub-band have no physical meaning. So,
magentic scatterers with parallel resonance are useless for
obtaining the left-handed medium. This makes unloaded long loops
be not prospective for this purpose.

\begin{figure}
\centering \epsfig{file=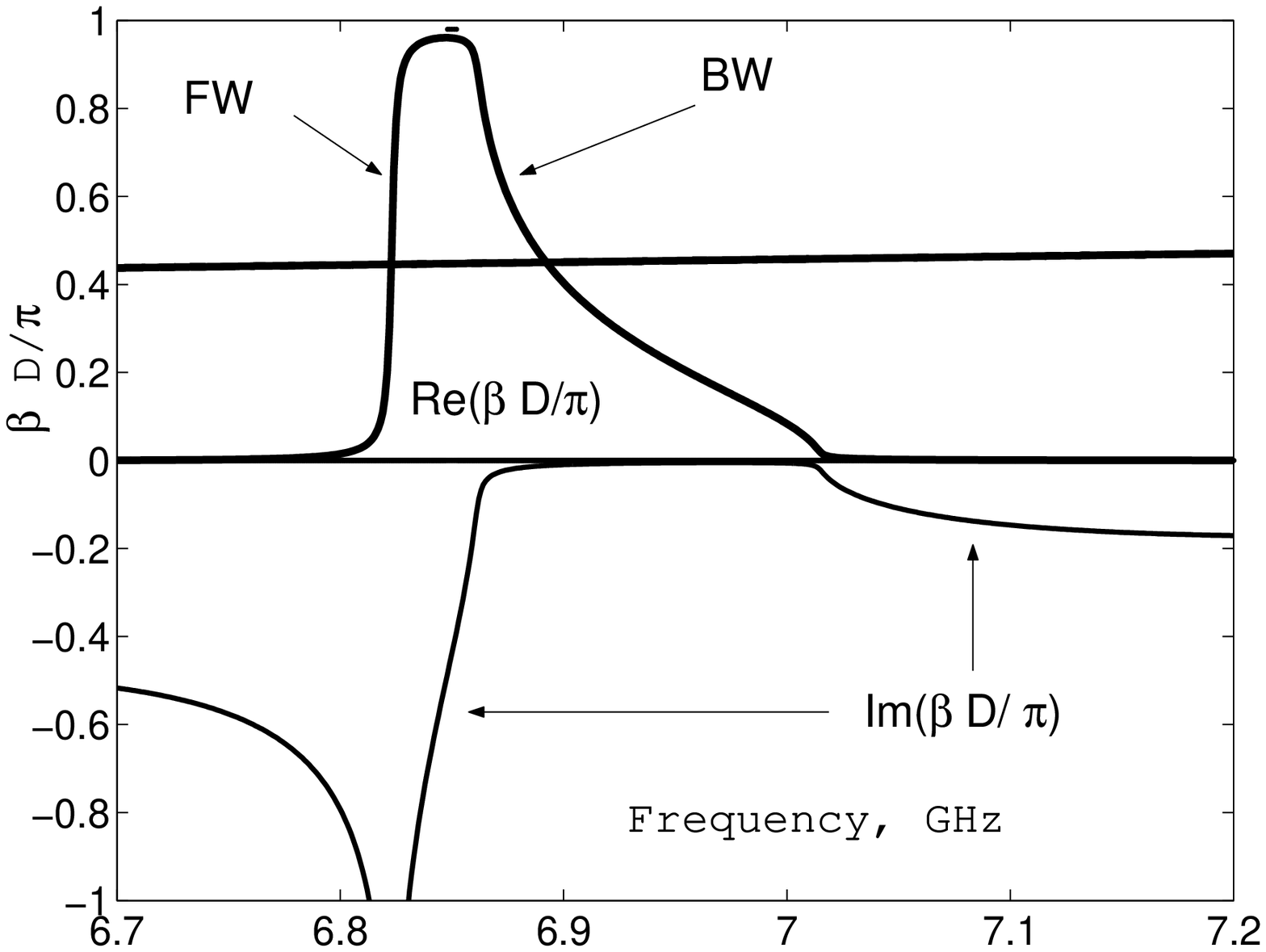, width=7.5cm}
\epsfig{file=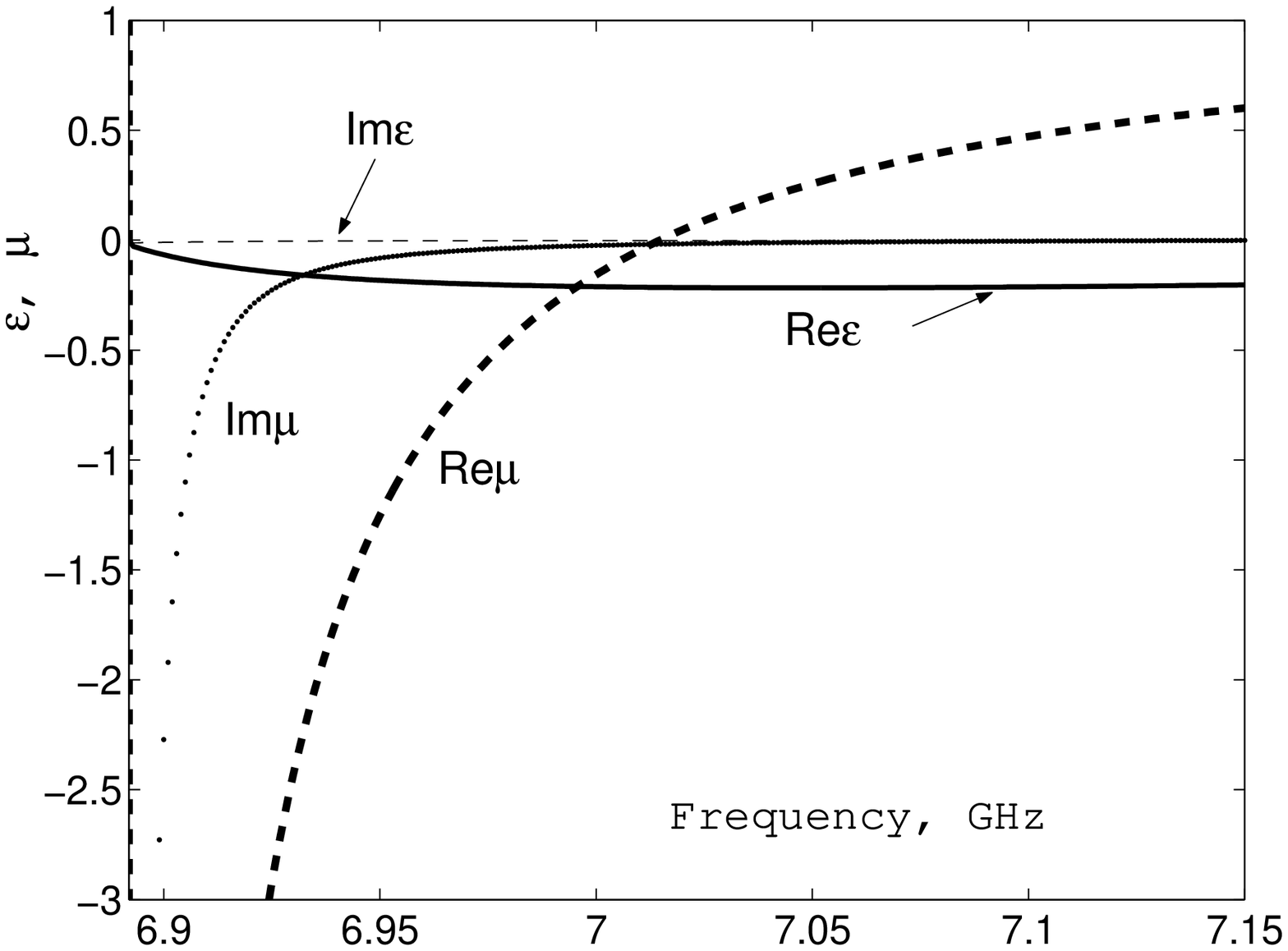, width=7.5cm}
 \caption {Left: dispersion plot within the resonant band of loaded
loops from copper wires. Forward wave (FW) is within the
lower-sub-band. Backward wave (BW) is within the upper sub-band.
Right: effective constitutive parameters $\epsilon_{xx}$ and
$\mu_yy$ (real and imaginary parts) within the band of the
backward wave.} \label{fig5}
\end{figure}

However, the case of loaded loops looks like very promising. In
Fig. \ref{fig5} the dispersion curves shown in the right part of
Fig. \ref{fig3} is reproduced. It is plotted within the second
resonant band of our MRs so that to see the resonant dispersion in
details. In the right part of the Figure one shows in details the
resonant behavior of effective constitutive parameters. One can
see that the lower sub-band of the series resonance band
($6.82-6.87$ GHz) corresponds to the forward wave propagating with
very strong attenuation. This wave substitutes the complex mode of
lossless lattice, and within this sub-band one has ${\rm
Re}(\mu_{yy})>0$ whereas ${\rm Re}(\epsilon_{xx})<0$ (see also
Fig. \ref{fig4}, right part). Within the upper sub-band of the
series resonance (from $6.87$ to $7.01$ GHz) one has the backward
wave regime. Within this band the imaginary parts of material
parameters do not exceed $4\%$ of their real parts. The band
$6.93-7.01$ GHz corresponds to values of the product
$|\sqrt{\epsilon_{xx}\mu_{yy}}|$ of the order unity (which is
promising for application of the layer from such a meta-material
as a lens). Imaginary parts of material parameters are very small
within this band ($1-2\%$).

At the first glance, one can think that the inductive loading of
long loops does not change their frequency properties. Unloaded
loops as well as loaded ones exhibit a set of parallel and series
resonances versus frequency. For $l=10$ mm we have the first
parallel resonance at $6$ GHz (see Fig. \ref{fig2}), the first
series resonance of unloaded loop holds at $12$ GHz, the second
parallel resonance at $16$ GHz, etc. To obtain a series resonance
at $6$ GHz with unloaded loops it is enough to take a loop with
$l=20$ mm. However, the calculations show that at this resonance
the structure with unloaded loops exhibits the very narrow
frequency band of the backward wave ($23$ MHz versus $124$ MHz
obtained above for loaded loops) and the magnetic losses for
unloaded loops are also higher than we obtained above for the case
of inductive loading of loops with $2l=17$ mm. The loading makes
the frequency positions of parallel and series resonances not
equidistant, broadens the resonant band and decreases the losses
in the left-handed regime.

\subsection*{4.~Conclusion}

We have suggested a new variant of a uniaxial left-handed material
for microwave applications which theoretically exhibits very small
losses in a comparatively wide frequency band ($2\%$). The
comparison with the two-phase lattice of straight wires and
split-ring resonators exhibits the better characteristics of the
proposed structure. This is the result of substituting the simple
circuit resonance of capacitively loaded small magnetic resonators
(SRRs) by the resonance of a loaded transmission line. Our
magnetic resonators are long and narrow inductively loaded loops.
In the present paper we suggest only an approximate analytical
model, which is, however, self-consistent and satisfying the basic
physical conditions. The further numerical simulations are
necessary. The material can be realized using the planar
technology which makes possible its experimental testing in the
next future.

%%% end here with text of summary

%%% Acknowledgements
%\vskip0.5cm
%{\noindent \bf Acknowledgement}\\
%The authors acknowledge the preparation of summaries according to these instructions.

%%% References
\end{document}